\documentclass{optica-article}

\journal{opticajournal} 

\articletype{Research Article}

\begin{document}

\title{Low-Complexity Monitoring and Compensation of Transceiver IQ Imbalance by Multi-dimensional Architecture for Dual-Polarization 16 Quadrature Amplitude Modulation }

\author{Yukun Zhang,\authormark{1},Xiaoxue Gong,\authormark{1,*},Xu Zhang\authormark{2} and Lei Guo\authormark{2}}

\address{\authormark{1}School of Communications and Information Engineering, Chongqing University of Posts and Telecommunications, Chongqing 400065, China\\
\authormark{2}School of Computer Science and Engineering, Northeastern University, Shenyang 110167, China\\}

\email{\authormark{*}Gongxx@cqupt.edu.cn}

\begin{abstract*} 
In this paper, a low-complexity multi-dimensional architecture for IQ imbalance compensation is proposed, which reduces the effects of in-phase (I) and quadrature (Q) imbalance. The architecture use a transceiver IQ skew estimation structure to compensate for IQ skew, and then use a low-complexity MIMO equalizer to compensate for IQ amplitude/phase imbalance. In the transceiver IQ skew estimation structure, the receiver(RX) IQ skew is estimated by Gardner’s phase detector, and the transmitter TX skew is estimated by finding the value that yields the lowest equalizer error.  The low-complexity MIMO equalizer consists of a complex-valued MIMO (CV-MIMO) and a two-layer multimodulus algorithm real-valued MIMO (TMMA-RV-MIMO), which employ a butterfly and a non-butterfly structure, respectively. The CV-MIMO is used to perform polarization demultiplexing and the TMMA-RV-MIMO equalizes each of the two polarizations. In addition, the TMMA-RV-MIMO can recovery the carrier phase. A 100 km transmission simulation and experiment with 36 Gbaud dual-polarization 16 quadrature amplitude modulation (DP-16QAM) signals showed that, with the TX/RX IQ skew estimation, the estimation error is less than 0.9/0.25 ps. The low-complexity MIMO equalizer can tolerate $\pm$0.1 TX IQ amplitude imbalance and $\pm$5 degrees at a 0.3 dB Q-factor penalty. The number of real multiplications is reduced by 55\% compared with conventional cases in total.

\end{abstract*}

\section{Introduction}

 To accommodate the increasing demand for data rates, optical communication systems are transitioning to higher symbol rates and high-spectral-efficiency modulation formats. High-order quadrature amplitude modulation (QAM) with a high baud rate has the potential to increase the data rate and spectral efficiency. However, the QAM signal requires a complex transceiver to send and receive. For example, the dual-polarization optical QAM signal passes through a dual-polarization $90^\circ$ optical hybrid, a balanced photodetector(BPD) and a transimpedance amplifier(TIA) to be demodulated. These components may impair the signal under non-ideal conditions. In a non-ideal condition, it is common to have a timing misalignment (IQ skew) of several picoseconds between the in-phase (I) and quadrature (Q) paths, or amplitude/phase imbalance between I and Q. These impairments are caused by waveguide length mismatches between I/Q arms, TIA delay and gain imbalance, or other imperfections in the optical front-end. These imbalances are commonly calibrated during the manufacturing process. However, some residual imbalances can remain. In addition, environmental and mechanical factors, including temperature fluctuations and device aging, cause dynamic changes in IQ imperfection. At higher symbol rates, this post-calibration residual imbalance has a more severe impact than on lower symbol-rate transmissions. The technical document for the 400ZR coherent optical interface stipulates that IQ skew should be less than 0.75 ps while IQ amplitude and phase imbalances must not exceed 1 dB and ±5° respectively. Thus, for high symbol-rate transmissions, an algorithm robust to IQ imbalance or a method to estimate the IQ imbalance is required.

Recently, several IQ imbalance-tolerant MIMO adaptive equalizers have been proposed. A well-known real-valued 4$\times$4 MIMO structure\cite{ref1} was proposed to compensate for IQ imbalance. However, the structure will fail when the CD is compensated for using a static equalizer\cite{ref2}. To overcome this problem, some new MIMO structures \cite{ref2,ref3,ref4,ref5,ref36} were proposed to compensate for IQ imbalance. However, the conventional receiver IQ imbalance compensation methods based on the AEQ suffer from IQ skew-enhanced timing jitter incurred by the clock recovery algorithm[9]. In addition, imbalance-tolerant MIMO equalizers require more taps and multiple iterations, which introduces an increase in complexity. Skew-tolerant MIMO demands more area in practical ASIC implementations in comparison to common complex-valued 2×2 MIMO equalizers.

Therefore, some IQ skew monitor algorithms based on a clock recovery algorithm were proposed. Reference \cite{ref11,ref12,ref13,ref18,ref35} use a digital Godard’s phase detector (Godard’s PD) to monitor the transceiver’s IQ skew. Reference \cite{ref14,ref15,ref16} propose receiver IQ skew tolarent scheme using a modified Gardner’s phase detector (Gardner’s PD) algorithm. In general, Gardner’s PD is better than Godard’s PD because Gardner’s PD doesn’t require the extra FFT. However, the IQ skew monitor based on Gardner’s PD algorithm is suitable only for short fiber transmission length due to fiber polarization mode dispersion (PMD) and chromatic dispersion (CD) compensation algorithm. When the signal polarization rotation (SOP) angle crosses 42.5$^\circ$ and the differential group delay (DGD) is 1/2 symbol period, the skew estimation will fail in Gardner’s PD algorithms \cite{ref11,ref18,ref17}. Furthermore, the skew monitor using Gardner’s PD is influenced by CD compensation. The reason is thar Rx IQ skew are added to a signal after accumulation of CD, but when CD compensation is performed first, the IQ skew become mixed in a complicated way. Some methods estimate the IQ imbalance based on data-aid. Reference \cite{ref6,ref7,ref8} are able to achieve wide-range IQ skew calibration simultaneously based training symbols.

  Because the IQ imbalance varies slowly, it is unnecessary to monitor it frequently. The current methods that compensate for the IQ imbalance are inefficient. For example, equalizers add more taps to compensate for receiver IQ imbalance. The methods that estimate the IQ imbalance based on data-aid need extra area in the circuit and increase the complexity. In addition, the IQ skew is the primary factor influencing the Q-factor in IQ imbalance. Reference.\cite{ref0} shows that the effective SNR reduction factor is IQ skew. Reference\cite{ref37} shows that for a 100 GBaud 16QAM signal, the IQ skew should be less than 1.1 ps when the signal-to-noise ratio(SNR) penalty is less than 1 dB. Therefore, the IQ skew should first be extracted using low-hardware-resource methods, and then equalizer can reduce the number of taps to equalize the signal. Skew monitoring based on Godard’s PD is a low-hardware-resource method, but it needs FFTs. 

 In this paper, we introduce a low-complexity IQ imbalance compensation architecture for DP-16QAM in 100 km single-mode fiber(SSMF) transmission. First, the transceiver IQ skew is estimated and compensated. We use Gardner’s PD to estimate the receiver IQ skew. Due to the low PMD coefficient, Gardner’s PD remains robust in 100 km transmission length. To avoid the IQ mixed, we set the CD compensation in the transmitter(TX) DSP. Moreover, for low complexity, we select the chirp filtering technique to compensate for CD. Then, we configure a scanning delay algorithm(SDA) to estimate the TX IQ skew. The SDA estimates and compensates the TX IQ skew by finding the value that yields the lowest equalizer error. Second, we propose a low-complexity equalizer to compensate for the TX amplitude and phase imbalance. We combine a complex-valued MIMO(CV-MIMO) and a TMMA-RV-MIMO for signal equalization, which employs a butterfly and a non-butterfly structure, respectively. The CV-MIMO is used to perform polarization demultiplexing. The TMMA-RV-MIMO equalizes each of the two polarizations and simultaneously compensates for the TX IQ imbalance. Since the TX and RX skew are being compensated, the new equalizer can equalize the signal with fewer taps.
For convenience, the new equalizer is called C-R AEQ. The 36 GBaud DP-16QAM simulation and experiment results demonstrate that the RX IQ skew estimation error is less than 0.25 ps, and the TX IQ skew estimation error is less than 0.9 ps. The C-R AEQ tolerates phase imbalance up to ±5 degrees and amplitude imbalance up to ±0.1 with a 0.3 dB Q-factor penalty. The number of real multiplications is reduced by 50\% compared with conventional cases in total.

\section{Working principle and simulations}
\subsection{Working principle}
The signal can be expressed as $ s(t)=s_I(t) + js_Q(t)$. Assuming that the signal $\hat{s}_1$(t) is added IQ skew by a delay $\tau $ in receiver, which can be expressed as  
\begin{equation}
\hat {s}_1(t) = s_I(t) \ast \delta\left(t-\frac{\tau}2\right)+j×s_Q(t) \ast \delta\left(t+\frac{\tau}2\right) 
\label{eq1}
\end{equation}

where the $\ast$ represents a convolution operation. Let $S(\omega)=\mathcal{F}\left\{\hat s_1(t)\right\}$., in frequency domain, the Eq.\ref{eq1} can be expressed as
\begin{equation}
\hat S(\omega)= S_I(\omega)e^{-j\omega\tau/2}+jS_Q(\omega)e^{j\omega\tau/2} \label{eq2}
\end{equation}
\begin{equation}
 S_{\omega} =\mathcal{F}\left\{s(t) \right\} 
\label{eq985}
\end{equation}
Where the $S_I(\omega)$ is $\Re(S_{\omega})$, and $S_Q(\omega)$ is $\Im(S_{\omega})$. According to Eq. \ref{eq2}, the delay $\tau$ introduces respective linear phases shift to $S_I(\omega)$ and $S_Q(\omega)$ in frequency domain. Therefore, the transfer function of delay $\tau$ in the frequency domain can be calculated as
\begin{equation}
H_\tau(\omega)=
     \begin{bmatrix}
e^{-j\omega\tau/2} & 0\\ 
0 & e^{j\omega\tau/2}\\
    \end{bmatrix}
\label{eq3}
\end{equation}

\begin{figure}[ht!]
\centering
\includegraphics[width=1\textwidth,keepaspectratio]{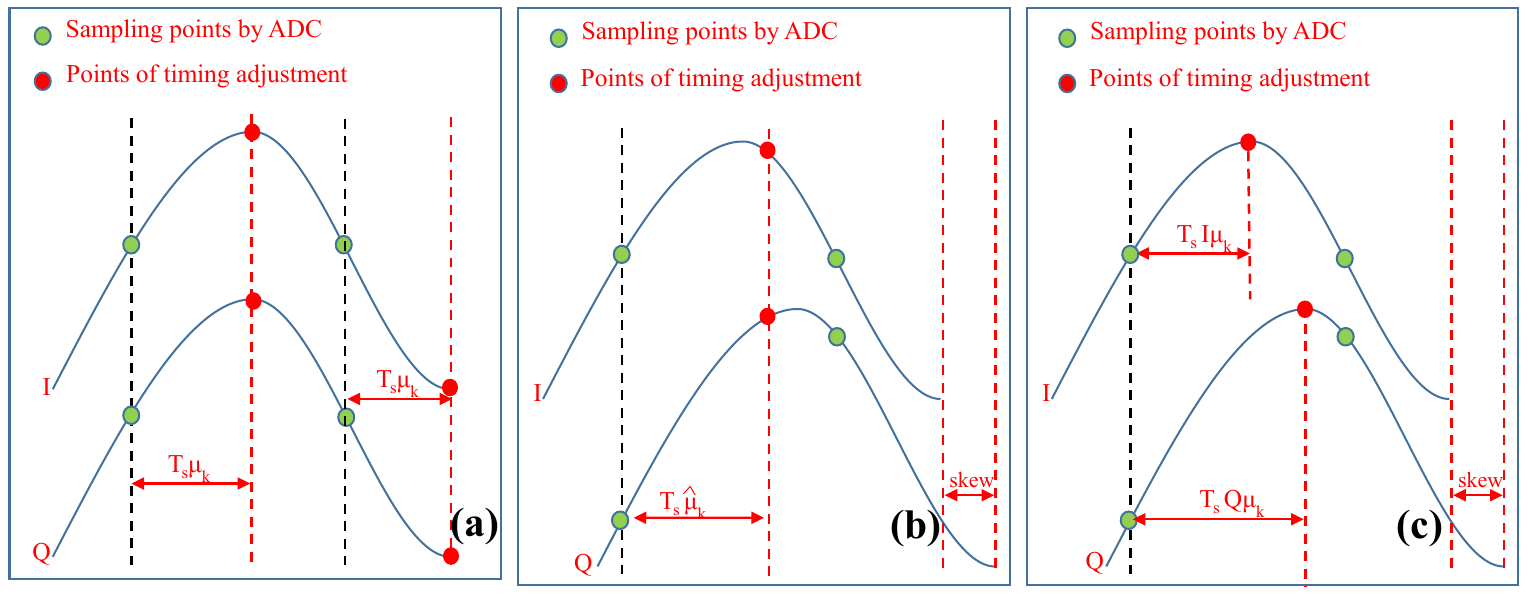} 
\captionsetup{justification=justified,singlelinecheck=false}
\caption{(a)Clock recovery Without IQ skew(b)Clock recovery With IQ skew,(c)Principle of IQ skew monitoring by clock recovery}
\label{Fig.1}
\end{figure}

  In general, this delay cannot be corrected by the all-digital feedback clock recovery algorithm(CRA). The CRA completes the timing adjustment by calculating the fractional interval $\mu_k$\cite{ref19}. When the clock is recovered, the sample point will be adjusted to the optimum detection of the symbol, as shown in Fig. \ref{Fig.1}(a).
 However, if there are IQ skews in X polarization or Y polarization, the numerically controlled oscillator(NCO) outputs $\hat\mu_k$ leading to the error sampling instant, as shown in Fig. \ref{Fig.1}(b). 
 So, if we calculate the I{$\mu_k$ and Q$\mu_k$  separately, as shown in Fig.\ref{Fig.1}(c), the IQ skew can be determined as
\begin{align}
X_{\tau x} &= \frac{1}{2×Baud}×mean(U_{XI}-U_{XQ})
\label{eq4}\\
Y_{\tau y} &= \frac{1}{2×Baud}×mean(U_{YI}-U_{YQ})        
\label{eq5}
\end{align}

Where $X_{\tau x}$, $Y_{\tau y}$ are the receiver IQ skew of the X polarization and Y polarization(pol), Baud is the symbol rate. 2Baud is the sample rate. The notation $\text{mean()}$ represents the average calculation.
$U_{XI}, U_{XQ}, U_{YI}, U_{YQ}$ are fractional intervals of I and Q in each Pol.

\begin{figure}[ht!]
\centering
\includegraphics[width=0.65\textwidth,keepaspectratio]{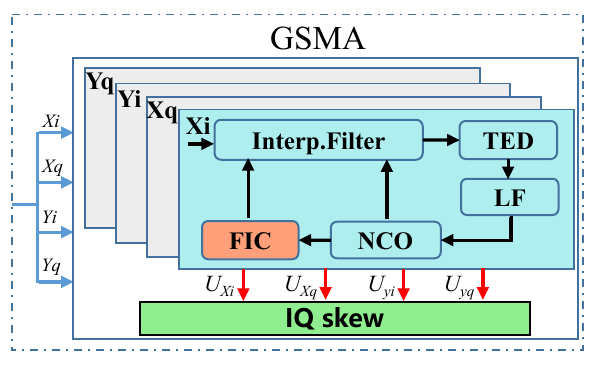} 
\captionsetup{justification=justified,singlelinecheck=false}
\caption{Diagram of receiver IQ skew monitoring}
\label{Fig.3}
\end{figure}
  Fig.\ref{Fig.3} illustrates Gardner’s skew monitoring algorithm(GSMA), deployed at the receiver, which employs Gardner’s PD as its timing error detector(TED). The algorithm inputs the I and Q components of both polarizations (Xi, Xq, Yi, Yq) into four separate CRAs. The outputs of the GSMA are the fractional interval $U_{XI}, U_{XQ}, U_{YI}$, and $U_{YQ}$. Using these values, the IQ skew for the corresponding polarization is calculated via Eq.\ref{eq4} and Eq.\ref{eq5}. 

\begin{figure}[ht!]
\centering
\includegraphics[width=1\textwidth,keepaspectratio]{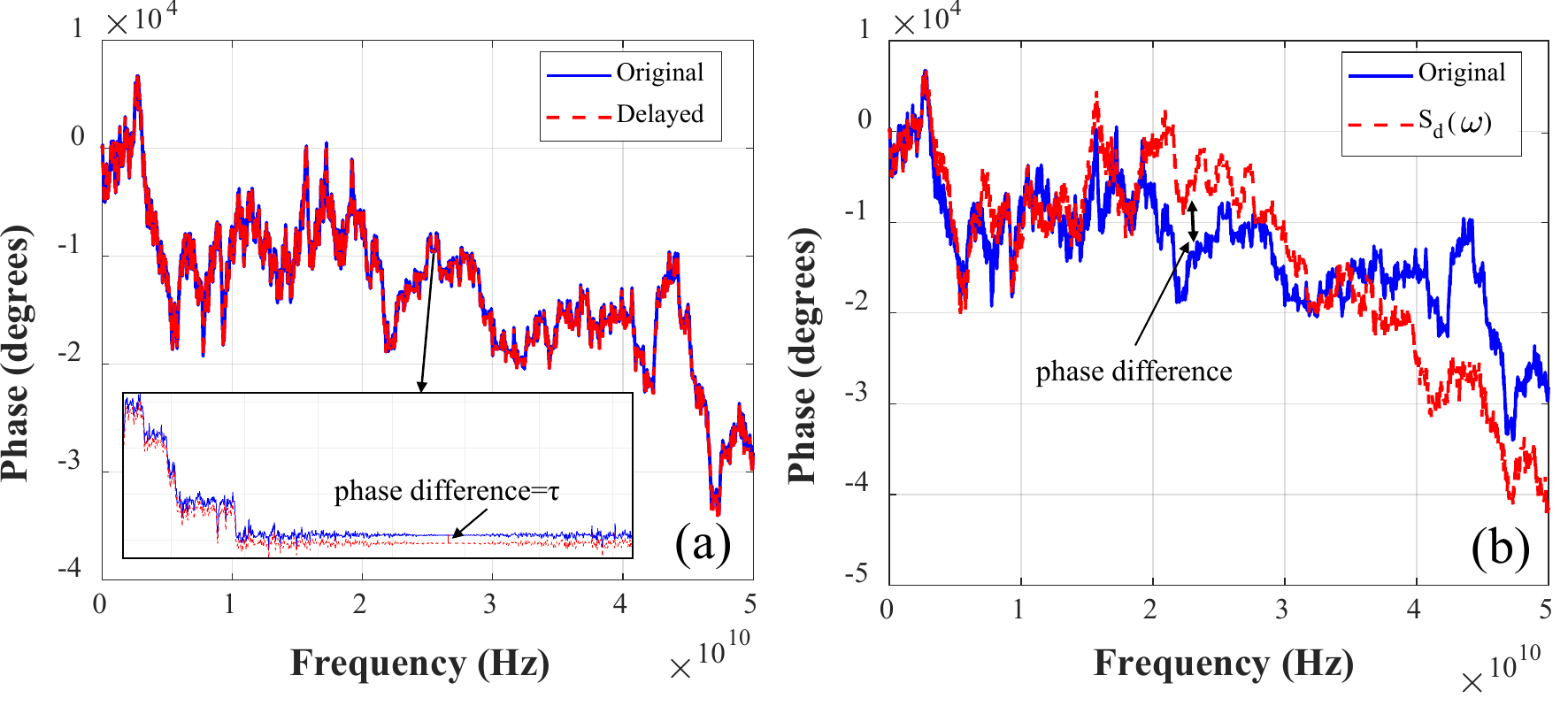} 
\captionsetup{justification=justified,singlelinecheck=false}
\caption{(a)The phase difference between the original and delayed signal without CD compensation(b)The phase difference between the original and delayed signal with CD compensation}
\label{Fig.2}
\end{figure}

 However, the effectiveness of the GSMA is impaired by CD compensation. When CD is compensated at the receiver, the algorithm fails to accurately estimate the RX IQ skew. This limitation is analyzed using the WL model in \cite{ref4}.
Let $h_{cd}(t) = \mathcal{F}^{-1}\left\{H_{cd}(\omega)\right\}$,we can write $h_{cd}(t)=h_{cd,I}(t)+j×h_{cd,Q}(t)$. Thus, the $\hat {s}(t)$, which suffers from CD and receiver IQ skew, can be represented as
\setlength{\arraycolsep}{1pt}
\begin{equation}
\begin{bmatrix}
\hat {s}_{I}(t)\\
\hat {s}_{Q}(t)\\
\end{bmatrix}
\hspace{0em} 
=
\hspace{0em} 
\begin{bmatrix}
h_{cd,I}(t) & -h_{cd,Q}(t)\\
h_{cd,Q}(t) & h_{cd,I}(t)\\
\end{bmatrix}
\hspace{0em} 
\begin{bmatrix}
\delta\left(t-\frac{\tau}2\right)  & 0\\
0 & \delta\left(t+\frac{\tau}2\right)\\
\end{bmatrix}
\hspace{0em} 
\begin{bmatrix}
s_{I}(t)\\
s_{Q}(t)
\end{bmatrix}
\label{eq6}
\end{equation}
\setlength{\arraycolsep}{5pt}

Therefore,$\hat {s}(t)$ can be expressed as:
\begin{align}
\hat{s}(t) &= h_1(t) \ast s(t) + h_2(t) \ast s^*(t) \label{eq7} \\
h_1(t) &= \frac{1}{2} \left\{
   \begin{aligned}
   &h_{cd,I}\left(t-\frac{\tau}{2}\right) + h_{cd,I}\left(t+\frac{\tau}{2}\right) \\
   &+ j \left[ h_{cd,Q}\left(t+\frac{\tau}{2}\right) + h_{cd,Q}\left(t-\frac{\tau}{2}\right) \right]
   \end{aligned}
\right\} \label{eq8}\\
h_2(t) &= \frac{1}{2} \left\{
   \begin{aligned}
   &h_{cd,I}\left(t-\frac{\tau}{2}\right) - h_{cd,I}\left(t+\frac{\tau}{2}\right) \\
   &+ j \left[ h_{cd,Q}\left(t+\frac{\tau}{2}\right) - h_{cd,Q}\left(t-\frac{\tau}{2}\right) \right]
   \end{aligned}
\right\} \label{eq9}
\end{align}

  Where $H_{cd}(\omega)$ is the frequency response of the CD. (.)$^*$ denotes the complex conjugate operation.
let $\hat S(\omega)=\mathcal{F}\left\{\hat s(t)\right\}, S(\omega)=\mathcal{F}\left\{s(t)\right\}$.Using $H_{cd}^{-1}(\omega)$ compensate $\hat S(\omega)$,the result is:
\begin{alignat}{2}
H_{cd}(\omega)&=e^{\frac{-j(\beta_{2}\omega_{c})}{2} \omega^2L} \label{eq10} \\
 S_d(\omega) &=H_{cd}^{-1}(\omega)×\hat S(\omega) \notag \\
     &= cos(\frac{(\omega×\tau)}{2})×S(\omega)-j×[H_{cd}^*(\omega)]^2×sin(\frac{(\omega×\tau)}{2})×S^*(-\omega) && \label{eq11}
\end{alignat}

  Where $S_{d}(\omega)$ is the signal after CD compensation in the frequency domain. The GSMA require estimate the phase shift $\tau$ from $S_{d}(\omega)$. However, as seen from Eq.\eqref{eq11}, $S_{d}(\omega)$ does not maintain a strictly linear phase. Consequently, the receiver IQ skew becomes undetectable when CD compensation is applied at the receiver.

This difference is illustrated in Fig.\eqref{Fig.2}. Fig. \eqref{Fig.2}(a) shows the phase shift between the delayed and original signals in the frequency domain. The delayed signal always keeps a fixed phase difference between the delayed signal and the original signal. Fig.\ref{Fig.2}(b) shows the phase shift between the $S_{d}(\omega)$ and the original signal in the frequency domain. It is observed in Fig.\eqref{Fig.2}(b) that the phase difference changes when the $\omega$ changes. Therefore, the delay $\tau$ cannot be detected by the GSMA.

  The IQ pha./amp imbalance is mixed in the same way. Assuming the signal is added amplitude and phase imbalance, which can be expressed as
\begin{align}
\begin{bmatrix}
s^{'}_{I}(t) \\ 
s^{'}_{Q}(t)\\
\end{bmatrix}
&=
\begin{bmatrix}
1  & 0\\
(1+g)sin(\theta)&(1-g)cos(\theta)\\
\end{bmatrix}
×
\begin{bmatrix}
s_{I}(t) \\ 
s_{Q}(t) \\
\end{bmatrix}
\end{align}

Where g is the amplitude imbalance, and $\theta$ is the phase imbalance. If the signal suffers from CD, which can be expressed as
\begin{equation}
\begin{bmatrix}
\hat {s}_{HI}(t)\\
\hat {s}_{HQ}(t)\\
\end{bmatrix}
\hspace{0em} 
=
\hspace{0em} 
\begin{bmatrix}
h_{cd,I}(t) & -h_{cd,Q}(t)\\
\substack {h_{cd,I}(t)(1+g)sin(\theta)-\\
h_{cd,Q}(t)(1-g)cos(\theta)} &\substack { h_{cd,Q}(t)(1+g)sin(\theta)+\\
h_{cd,I}(t)(1-g)sin(\theta)}\\
\end{bmatrix}
\hspace{0em}
\begin{bmatrix}
s_{I}(t)\\
s_{Q}(t)
\end{bmatrix}
\label{eq201}
\end{equation}
Similar to Eq. \eqref{eq11}, the $H_{CD}^{-1}(\omega)$ is used to compensate for CD, and the result can be expressed as
\begin{align}
\begin{aligned}
  S_H(\omega) &= (\frac{1}{2}+\frac{1}{2}e^{j\theta}+\frac{g}{2}(e^{-j\theta}))×S(\omega) \\
        &+(\frac{1}{2}-\frac{1}{2}e^{-j\theta}+\frac{g}{2}(e^{-j\theta})))×[H_{cd}^*(\omega)]^2×S^{*}(-\omega)
\label{eq202}
\end{aligned}
\end{align}

   $S_H(\omega)$ contains a component that consists of the complex conjugate of the signal. This component acts as interference that degrades the performance of the equalizer. According to Eq. \eqref{eq202}, it can be observed that the impact of this conjugate term on $S_{H}(\omega)$ is lower when the $\theta$ and g are smaller. Thus, the equalizer can partly compensate for the IQ pha./amp. imbalance when these imbalances are small.
\begin{figure}[ht!]
\centering
\includegraphics[width=1\textwidth,keepaspectratio]{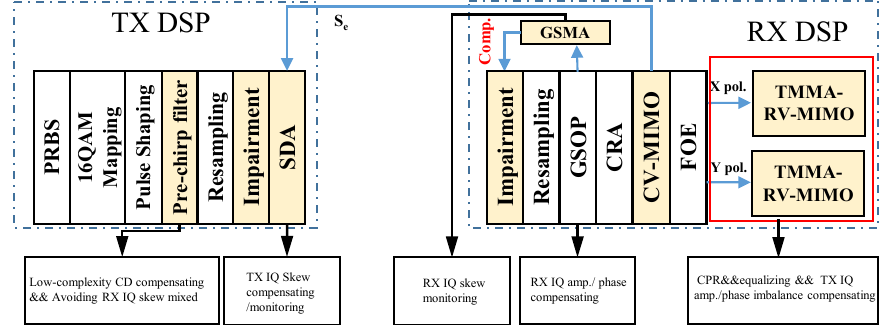} 
\captionsetup{justification=justified,singlelinecheck=false}
\caption{Proposed DPS flow of transceiver imbalance, SDA is scanning delay algorithm, GSMA is Gardner’s skew monitoring algorithm}
\label{Fig.5}
\end{figure}  

Based on the analysis above, we propose a DSP flow as shown in Fig. \ref{Fig.5}. In TX DSP, to address the issue that the receiver’s IQ skew cannot be extracted under large CD, we implement the CD compensation at the TX side. Furthermore, to reduce the complexity of CD compensation, we employ the chirp filter technique. The transfer function of the pre-chirp filter can be expressed as\cite{ref22}
\begin{align}
W &= e^{\frac{jT_{s}^2}{(2\beta_{2}L)}} \label{eq12} \\
y_{n} &= W^{n^2}FFT\left(x_kW^{k^2}\right)\label{eq13}
\end{align}
where $x_k$ is the input signal, $y_n$ is the output signal. $T_s$ is the sample period, $\beta_2$ is the dispersion coefficient, and L is the transmission length. FFT is the Fast Fourier Transform. The chirp-based CD compensation technique requires a single FFT computation, reducing the overall complexity to about 50\% of the typical FDE technique. 
\begin{figure}[ht!]
\centering
\includegraphics[width=0.6\textwidth,keepaspectratio]{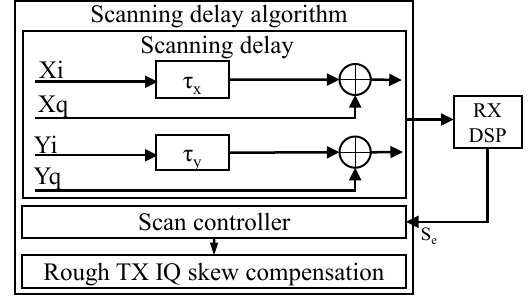} 
\captionsetup{justification=justified,singlelinecheck=false}
\caption{Diagram of SDA}
\label{Fig.scan}
\end{figure}  
The scanning delay algorithm(SDA) is employed to compensate for the TX IQ skew. Figure \ref{Fig.scan} is the diagram of the SDA, which comprises a scanning delay block, a scan controller block, and a TX IQ skew compensation block. Within the scanning delay block, the X pol and Y pol signals are subject to a gradually increasing delay from $-\hat nT/16$ to $\hat nT/16$, where $\hat n $ is an integer. This process can be expressed as 

\begin{equation}
\begin{bmatrix}
XI_{o} \\ 
 XQ_{o}\\
\end{bmatrix}
=
\begin{bmatrix}
\delta\left(t-\frac{\tau_{x}}2\right)  & 0\\
0 & \delta\left(t+\frac{\tau_{x}}2\right)\\
\end{bmatrix}
×
\begin{bmatrix}
XI_{i} \\ 
XQ_{i}\\
\end{bmatrix}
\label{eq14}
\end{equation}
\begin{equation}
\begin{bmatrix}
YI_{o} \\ 
 YQ_{o}\\
\end{bmatrix}
=
\begin{bmatrix}
\delta\left(t-\frac{\tau_{y}}2\right)  & 0\\
0 & \delta\left(t+\frac{\tau_{y}}2\right)\\
\end{bmatrix}
×
\begin{bmatrix}
YI_{i} \\ 
YQ_{i}\\ 
\end{bmatrix}
\label{eq15}
\end{equation}
  Where the $\tau_{x}$ or $\tau_{y}$ represent the IQ skew of X pol or Y pol, respectively. The $\tau_{x}$ and $\tau_{y}$ is gradually increasing delay from $-\hat nT/16$ to $\hat nT/16$. In the RX DSP, the  CV-MIMO module outputs the $S_{e}$ to the scan controller block after the scanning delay block. Subsequently, the scan controller block calculates the error $S_{e}$. The $S_{e}$ can be expressed as 
\begin{align}
S_{e}&=\frac{E_{\tau_{x}}+E_{\tau_{y}}}{2}
\label{eq158}\\
E_{\tau_{x}}&=\sum_{n=1}^{N}\frac{e(n)_{x}}{N}
\label{eq16}\\
E_{\tau_{y}}&=\sum_{n=1}^{N}\frac{e(n)_{y}}{N}
\label{eq17}
\end{align}
Where N is the number of symbols, $e(n)_{x}$/$e(n)_{y}$ is X/Y pol equalization error. During the scanning process, the scan controller block records a set of $S_{e}$, and identifies the minimum $S_{e}$. Then, the $\tau_{x}$ and $\tau_{y}$ that correspond to this minimum $S_e$ are output as the estimated TX skew. Subsequently, the TX IQ skew compensation block uses these estimated delays to compensate for the TX IQ skew. The RX IQ pha./amp. imbalance is compensated for using GSOP. The RX IQ skew is compensated for using GSMA. The GSMA diagram is shown in Fig. \ref{Fig.3}.
\begin{figure}[ht!]
\centering
\includegraphics[width=0.5\textwidth,keepaspectratio]{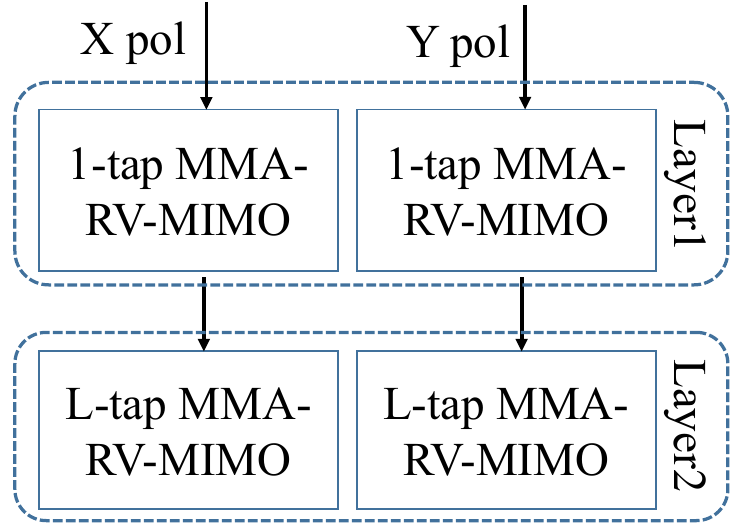} 
\captionsetup{justification=justified,singlelinecheck=false}
\caption{Diagram of TMMA-RV-MIMO}
\label{Fig.68}
\end{figure}
 To compensate for the TX IQ amp./pha. imbalance and equalize signal with low complexity, we propose a C-R AEQ inspired by the KM AEQ [24]. The C-R AEQ is composed of CV-MIMO and TMMA-RV-MIMO. The CV-MIMO is composed of a butterfly structure with fewer taps per filter, as we found that frequency offset estimation (FOE) can be performed correctly after the signal is equalised by fewer taps CV-MIMO(such as 3-tap). The TMMA-RV-MIMO is a non-butterfly filter after the FOE as shown in Fig. \eqref{Fig.68}. 
The CV-MIMO is used for polarization demultiplexing and to ensure that FOE is correctly performed. TMMA-RV-MIMO equalizes each of the two polarisations and compensates for the TX IQ imbalance. Moreover, due to the phase sensitivity of TMMA-RV-MIMO, the TMMA-RV-MIMO can recover the carrier phase(CPR). The tap coefficient vector of the TMMA-RV-MIMO is updated according to ref. \cite{ref1}. C-R AEQ is similar to ref.\cite{ref26}. The difference is that ref.\cite{ref26} does not consider CD and frequency offset. This means ref.\cite{ref26} is only suitable for a self-homodyne coherent optical system in a short transmission length. And DD-LMS requires symbol decision , whereas C-R AEQ does not require it. In addition, ref.\cite{ref26} requires higher complexity.
\subsection{Simulation}
\begin{figure}[ht!]
\centering
\includegraphics[width=0.8\textwidth,keepaspectratio]{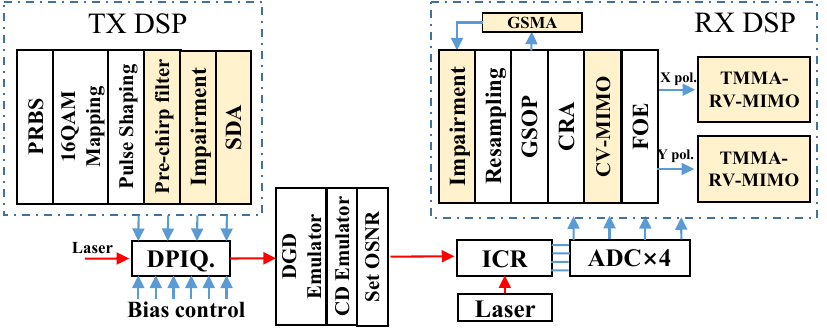} 
\captionsetup{justification=justified,singlelinecheck=false}
\caption{System setup for 36 GBaud 16QAM}
\label{Fig.50}
\end{figure}
  To confirm our proposed architecture, a 36 Gbaud dual-polarization Nyquist 16QAM (DP-Nyquist-16QAM) signal with the system diagram displayed in Fig.\eqref{Fig.50}. In the transmitter, the full raised cosine filter is used to implement the pulse shape, and the roll-off factor is set to 0.5. The samples per symbol(SPS) are 4, and the pre-chirp filter block size is 264. Transmit a total of 2$^{16}$ symbols. The TX skew is set from -6 ps to 6 ps (from -T/4 to T/4 of 36 Gbaud). We analyze the impact of polarization effects using the higher-order PMD model. The mean DGD is set to 0.5 ps/km$^{1/2}$ in the DGD emulator. The CD coefficient is $-16 ps/(nm \cdot km)$ in the CD emulator. The transmission length is 100 km. The OSNR is defined assuming a bandwidth of 12.5 GHz and is set to 23 dB. In the receiver, the RX skew is set from -6 ps to 6 ps. The signal is down-sampled to 2 SPS. The CV-MIMO tap numbers are set to 6 taps. The TMMA-RV-MIMO tap number is set to 22 taps. The transceiver imbalances are digitally added in the DSP model of Rx and Tx. The transmission laser and the LO laser are operating at 1552.5 nm with a 100 kHz linewidth and a 1 GHz frequency offset.
\subsubsection{The Effect of pre-CD compensation on RX IQ Skew Estimation}
\begin{figure}[ht!]
\centering
\includegraphics[width=0.65\textwidth,keepaspectratio]{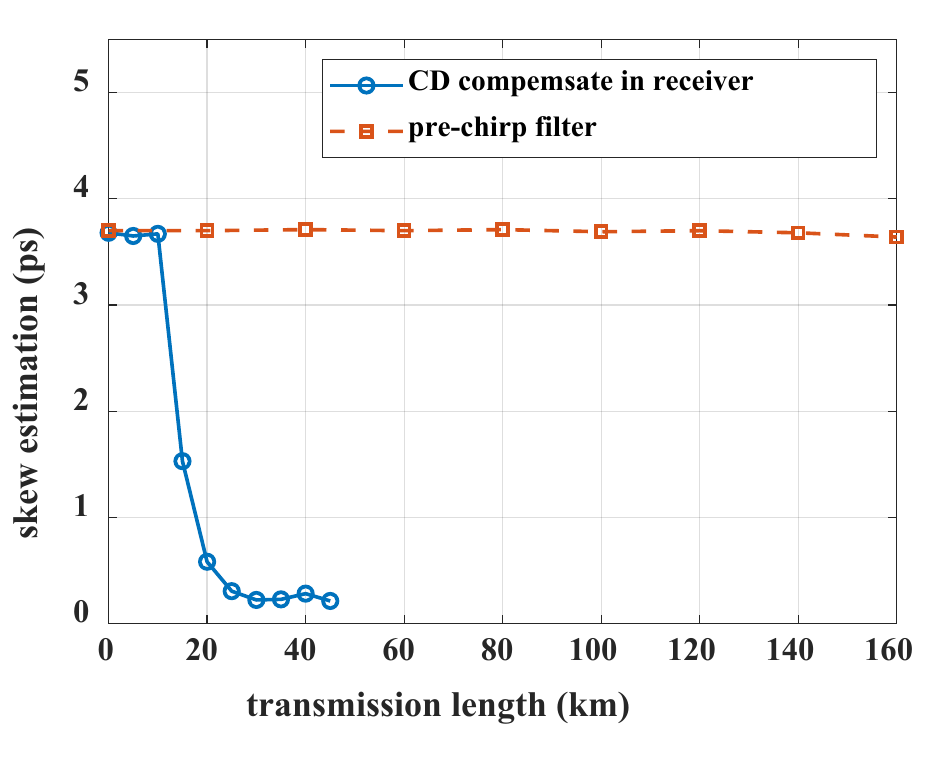} 
\captionsetup{justification=justified,singlelinecheck=false}
\caption{The Effect of pre-CD compensation on RX IQ Skew Estimation}  
\label{Fig.68}
\end{figure}
  First, the relation between the CD compensation and the RX IQ skew estimation is tested in simulation. In Fig. \ref{Fig.68}, the estimation results are plotted as a function of fiber length. To clearly observe the relationship, the RX X pol IQ skew is set at 3.9 ps, and other values of transceiver IQ imbalance are set to zero in this part. We use the pre-chirp filter in the transmitter or the FDE technique in the receiver to compensate for CD. It should be noted that the transmission length is less than 10 km, the CD compensation is not performed. The RX skew estimation is shown in Fig. \ref{Fig.68}. According to Fig. \ref{Fig.68}, it can be found that if the CD compensation is in the receiver, the RX IQ skew estimation is close to 0 ps when the transmitter length is larger than 20 km. But if the CD compensation is in the transmitter, the RX IQ skew estimation is close to 3.9 ps, and the result is unaffected by the transmission length. In general, receiver IQ skew is uncorrelated with the transmission length. However, from Fig. \ref{Fig.68}, if the optical signal has a larger CD and the CD is compensated in the receiver, the receiver IQ skew cannot be detected by GSMA. This simulation result demonstrates our proposed theory that the location of CD compensation is related to RX skew estimation, and the RX IQ skew can be detected after pre-CD compensation. 

\subsubsection{The Effect of Transceiver IQ Imbalance on RX IQ Skew Estimation}
\begin{figure}[ht!]
\centering
\includegraphics[width=1\textwidth,keepaspectratio]{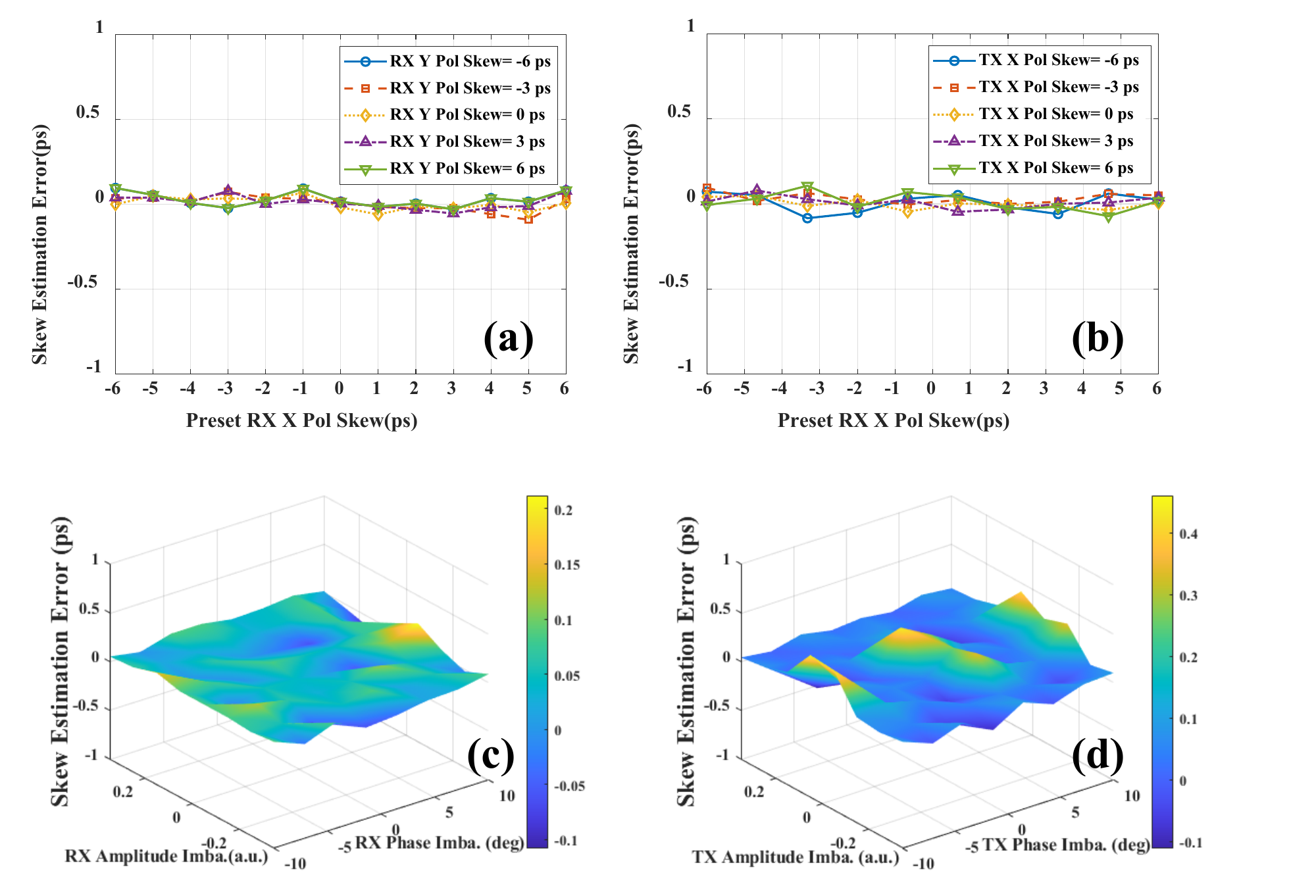} 
\captionsetup{justification=justified,singlelinecheck=false}
\caption{(a)The relation between skew estimate error and RX X/Y pol IQ skew (b)The relation between skew estimate error and TX X pol IQ skew(c) The relation between skew estimate error and RX pmp./pha. imbalance(d) The relation between skew estimate error and TX pmp./pha. imbalance}
\label{Fig.9}
\end{figure}
Next, the effect of the transceiver IQ imbalance on the RX IQ skew estimation is simulated. The RX X pol skew is set from -6 ps to 6 ps. In order to comprehensively observe the imperfect factor influence on the RX X pol skew estimation error, we consider the RX X pol amplitude imbalance, phase imbalance, RX Y pol skew, and TX X pol amplitude imbalance, phase imbalance, TX X pol skew. These results are shown via 2-Dimensional(2D) planes or 3-Dimensional(3D) surfaces. Unless otherwise stated, the basic IQ imbalance is X pol IQ imbalance. For example, RX skew is RX X pol skew, TX amplitude imbalance is TX X pol amplitude imbalance, RX skew estimation error is  RX X pol skew estimation error, and so on. The Tx and RX amplitude imbalance varies from -0.35 to 0.35. The TX and RX phase imbalance varies from -10 degrees to 10 degrees. The TX skew and RX Y pol skew varies from -6 ps to 6 ps, and the RX skew changes from -6 ps to 6 ps. Based on these conditions, the RX skew estimation error is shown in Fig.\ref{Fig.9}. The Fig.\ref{Fig.9}(a) displays the relation between RX skew estimation and RX Y pol skew. As the RX Y pol skew changes -6 ps to 6 ps, the RX skew estimation error is always kept less than 0.25 ps. This means that the RX skew estimation is unaffected by RX Y pol skew. The Fig.\ref{Fig.9}(b) displays the relation between RX skew estimation and TX skew. Due to the frequency offset isolate TX skew and RX skew[7], as the TX skew changes -6 ps to 6 ps, the RX skew estimation error is always kept less than 0.3 ps. This means that the RX skew estimation is unaffected by TX skew. The reason is that the frequency offset and phase noise isolate the TX imbalance. The Fig.\ref{Fig.9}(c) displays the relation between RX skew estimation, RX amplitude imbalance, and RX phase imbalance. According to Fig.\ref{Fig.9}(c), the RX skew estimation error is kept less than 0.25 ps when the amplitude imbalance changes from -0.35 to 0.35, and the phase imbalance varies from -10 degrees to 10 degrees. The Fig.\ref{Fig.9}(d) displays the relation between RX skew estimation, TX amplitude imbalance, and TX phase imbalance. According to Fig.\ref{Fig.9}(d), the RX skew estimation error is kept less than 0.5 ps when the amplitude imbalance changes from -0.35 to 0.35, and the phase imbalance varies from -10 degrees to 10 degrees. Therefore, according to  Fig.\ref{Fig.9}(a) - Fig.\ref{Fig.9}(d), it can be concluded that the RX skew estimation is unaffected by transceiver IQ imbalance.

\subsubsection{The Effect of Transceiver IQ Imbalance and OSNR on TX IQ Skew Estimation}
\begin{figure}[ht!]
\centering
\includegraphics[width=0.65\textwidth,keepaspectratio]{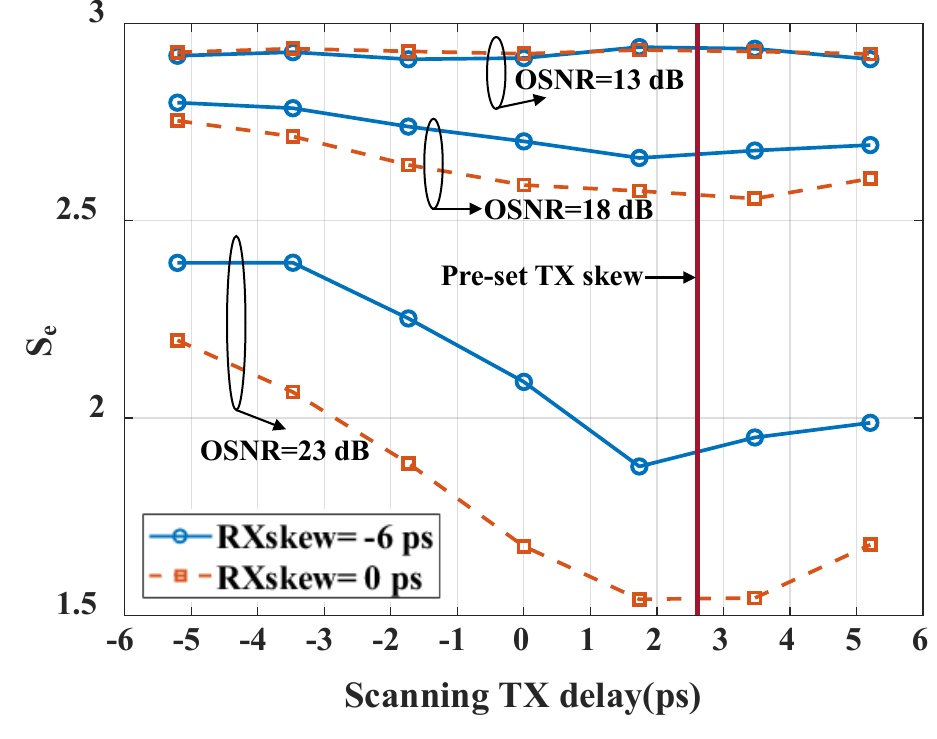} 
\captionsetup{justification=justified,singlelinecheck=false}
\caption{$S_{e}$ varies with the  OSNR and RX skew change}
\label{Fig.10}
\end{figure}
The TX IQ skew estimation is via the $S_{e}$ output by CV-MIMO. However, the error is affected by many factors. We mainly consider the effect of transceiver IQ imbalance and OSNR on TX IQ skew estimation. The OSNR is set to 13 dB, 18 dB, and 23 dB, respectively. The TX skew is fixed as 2.6 ps (1.5T/16 of 36 GBaud), and SDA scans the TX delay at T/16 intervals. The Tx and RX amplitude imbalance varies from -0.35 to 0.35. The TX and RX phase imbalance varies from -10 to 10 degrees. The RX skew is set to -6 ps, 0 ps, respectively. The TX Y pol skew varies from -6 to 6 ps. Fig.\ref{Fig.10} displays the relation between OSNR and TX IQ skew estimation. At OSNRs of 18 dB and 23 dB, $S_{e}$ is minimum at 1.74 ps (T/16 of 36 GBaud) or 3.47 ps (2T/16 of 36 GBaud). From the scan control rule, the TX skew estimation is 1.74 ps or  3.47 ps. However, at the OSNR of 13 dB, the $S_{e}$ can not display the correct TX skew estimation. The reason is that the noise is the main influence on $S_{e}$ instead of TX skew.

\begin{figure}[ht!]
\centering
\includegraphics[width=0.85\textwidth,keepaspectratio]{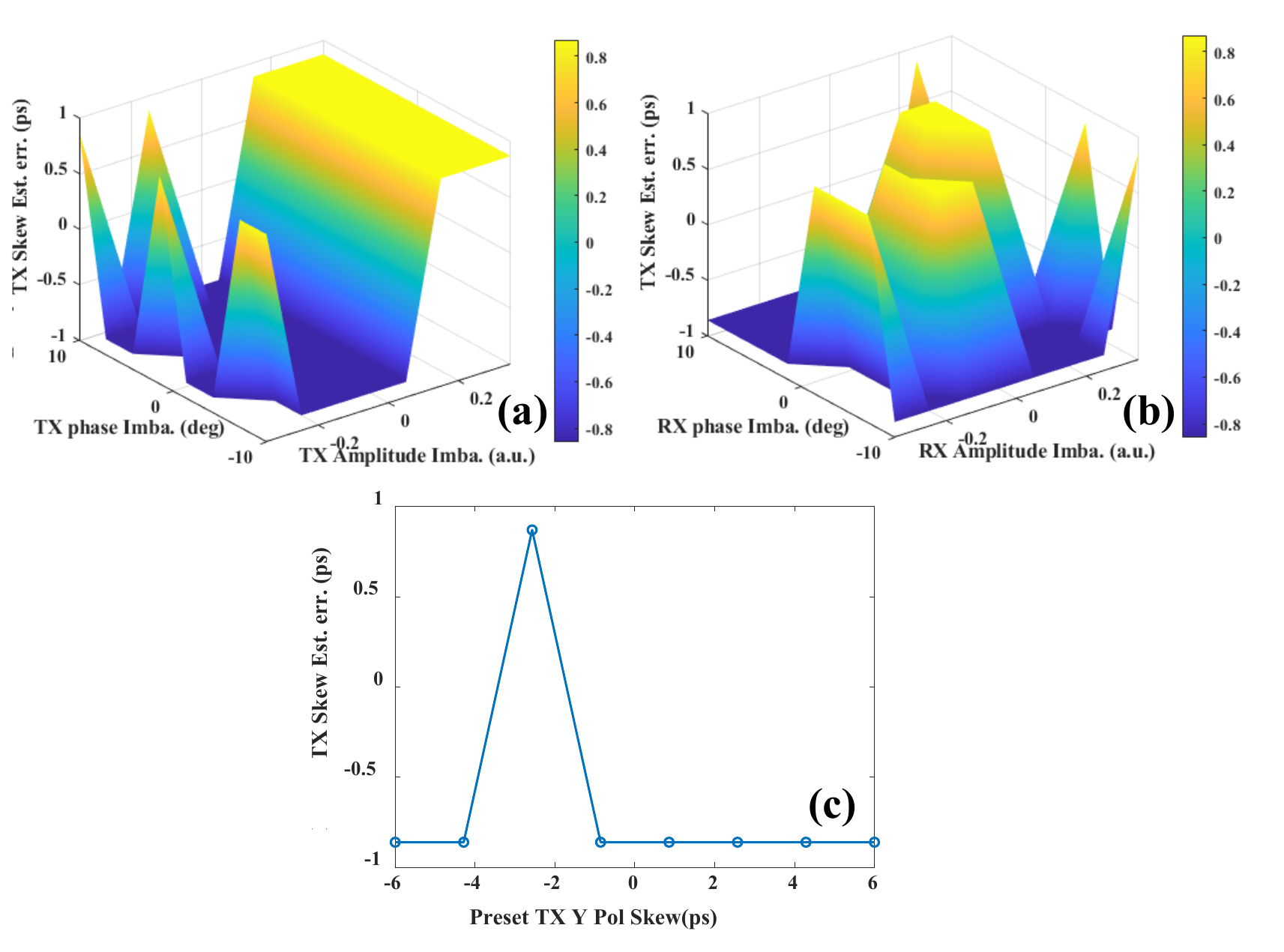} 
\captionsetup{justification=justified,singlelinecheck=false}
\caption{(a)TX skew estimation error with TX amp./pha. imbalance changing(b)TX skew estimation error with RX amp./pha. imbalance changing(c)TX skew estimation error  with TX Y pol change}
\label{Fig.11}
\end{figure}
Fig.\ref{Fig.11} displays the effect of transceiver IQ imbalance on TX IQ skew estimation. Fig.\ref{Fig.11}(a) and Fig.\ref{Fig.11}(b) display that the estimation error is kept less than $\pm$0.9 ps when the TX and RX phase and amplitude imbalance are changed. Fig.\ref{Fig.11}(c) displays that the estimation error is kept less than $\pm$0.9 ps when TX Y pol skew changes from -6 ps to 6 ps. According to Fig.\ref{Fig.11}, it is concluded that the TX skew estimation error is unaffected by transceiver IQ imbalance.

  It should be noted that transmitted X pol can be received as X or Y pol and vice versa. Therefore, simultaneous scan delay on both X and Y polarizations is not allowed. 
\subsubsection{Performance with Transmitter IQ Imbalance after C-R AEQ}
\begin{figure}[ht!]
\centering
\includegraphics[width=0.85\textwidth,keepaspectratio]{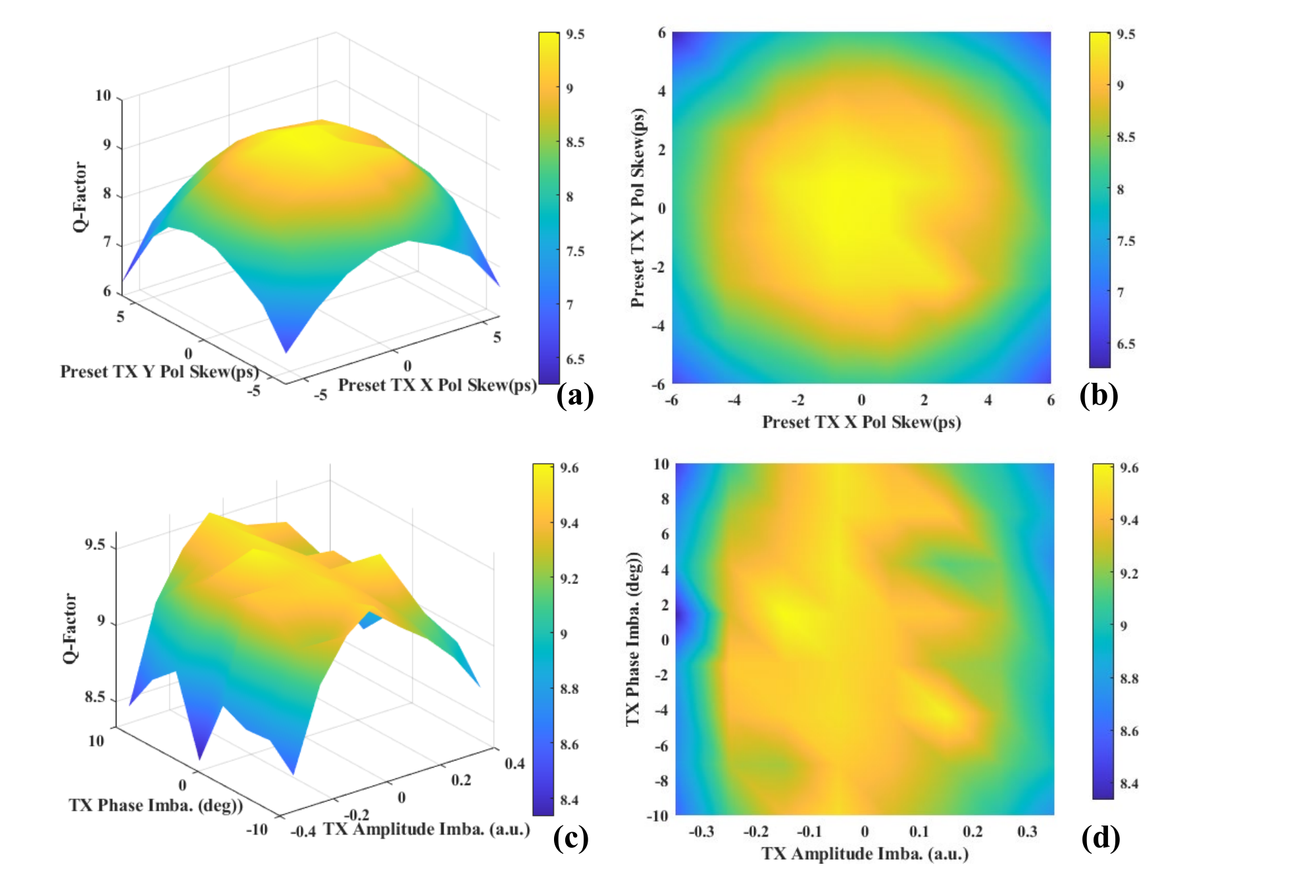} 
\captionsetup{justification=justified,singlelinecheck=false}
\caption{(a)TX skew tolerance of the proposed architecture after 100 km transmission(b) Project the 3D surface of (a) on TX X pol skew and TX Y pol skew (c)TX amp./pha. imbalance tolerance of the C-R AEQ after 100 km transmission(d) Project the 3D surface of (c) on TX amp. imbalance and TX pha. imbalance}
\label{Fig.12}
\end{figure}
   In this part, we assess the performance of C-R AEQ under the transmitter IQ imbalance condition. The Tx amplitude imbalance varies from -0.35 to 0.35. The TX phase imbalance varies from -10 degrees to 10 degrees. The TX skew and TX Y pol skew vary from -6 ps to 6 ps. The Q-factor is used to display the signal quality after C-R AEQ. In order to clearly observe which is the major factor influencing the Q-factor, we study not only the 3D surfaces but also their 2D projections. Fig.\ref{Fig.12}(a) displays the TX skew tolerance of C-R AEQ, and Fig.\ref{Fig.12}(b) is the 2D projections of Fig.\ref{Fig.12}(a). According to Fig.\ref{Fig.12}(a), the maximum Q is 9.5 dB when TX skew and TX Y pol skew are 0 ps. As the TX skew approaches $\pm$6 ps when the TX Y pol skew approaches $\pm$6 ps, the Q-factor is reduced to 6.3. This means that the TX skew is one of the major influences on the Q-factor. According to Fig.\ref{Fig.12}(b), the Q penalty is 0.15 dB when the Tx skew and TX Y pol skew are less than $\pm$2 ps($\pm$1/16 of symbol period). From Fig.\ref{Fig.10} and Fig.\ref{Fig.11}, the SDA makes the residual TX skew less than 1 ps. Fig.\ref{Fig.12}(c) shows the TX amplitude and phase imbalance tolerance of C-R AEQ, and Fig.\ref{Fig.12}(d) is the 2D projections of Fig.\ref{Fig.12}(c). Seen from Fig.\ref{Fig.12}(c), compare to Fig.\ref{Fig.12}(a), TX amplitude and phase imbalance influence on Q-factor is lower than TX skew. The Q-factor is 8.4 when the TX amplitude approaches $\pm$0.35, and the TX phase approaches $\pm$10 degrees. As shown in Fig.\ref{Fig.12}(d), when the TX phase imbalance approaches $\pm$10 degrees, and there is no TX amplitude imbalance, the Q-factor remains 9.5 dB. In addition, when the TX amplitude imbalance changes from 0 to $\pm$0.35, the Q-factor changes from 9.5 to 8.4 dB. This means that the TX phase imbalance has the least influence on Q-factor after C-R AEQ, whereas the amplitude imbalance has a greater influence on Q-factor than the phase imbalance. 

\subsubsection{Performance with Receiver IQ Imbalance after C-R AEQ}
\begin{figure}[ht!]
\centering
\includegraphics[width=0.85\textwidth,keepaspectratio]{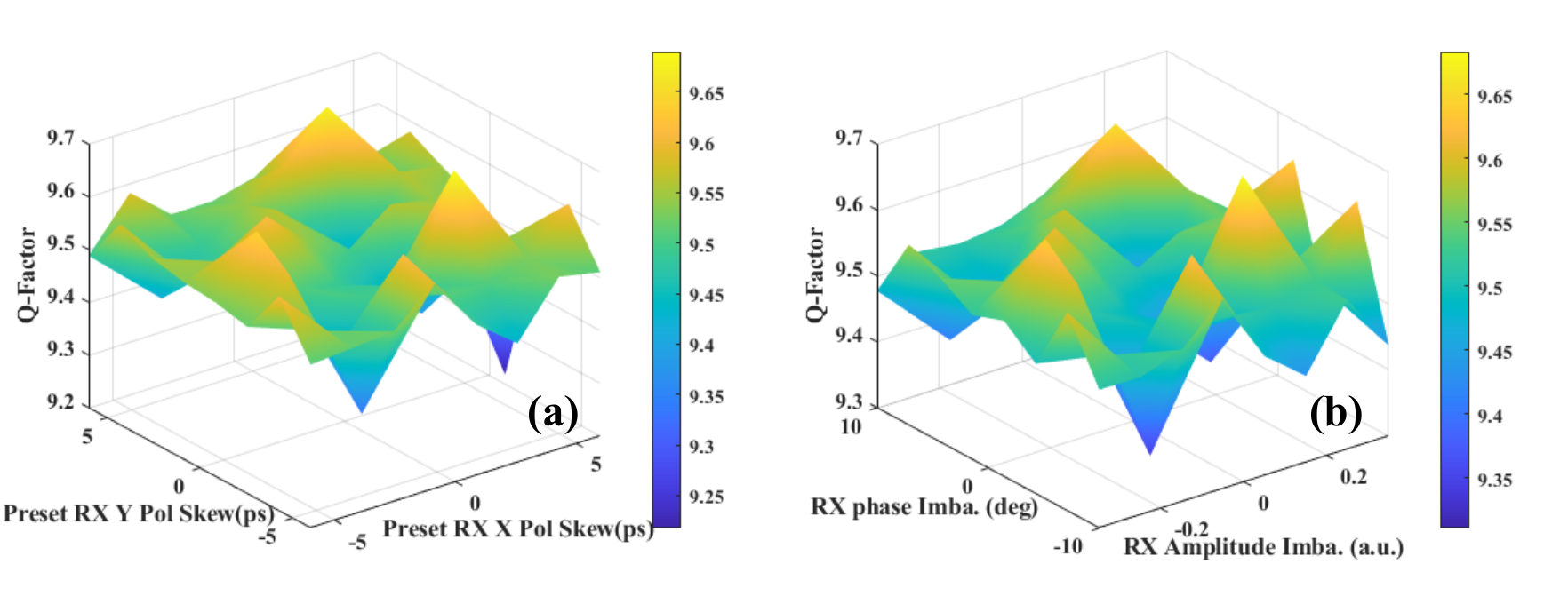} 
\captionsetup{justification=justified,singlelinecheck=false}
\caption{(a)RX skew tolerance of proposed architecture after 100 km transmission(b)RX amp./pha. imbalance tolerance of proposed architecture after 100 km transmission}
\label{Fig.13}
\end{figure}
In this part, we assess the performance of C-R AEQ under the receiver IQ imbalance condition. The Rx amplitude imbalance varies from -0.35 to 0.35. The RX phase imbalance varies from -10 degrees to 10 degrees. The RX skew and RX Y pol skew vary from -6 ps to 6 ps.Fig.\ref{Fig.13}(a) and Fig.\ref{Fig.13}(b) display the Q-factor performance with RX skew, Y pol skew, RX phase, and amplitude imbalance.
From  Fig.\ref{Fig.13}(a) and Fig.\ref{Fig.13}(b), the maximum Q-factor is greater than 9.65 dB, and the minimum Q-factor is greater than 9.25 dB. Therefore, due to the GSOP and GSMA, the Q-factor is unaffected by the RX IQ imbalance.
\subsubsection{Performance with Transceiver IQ Imbalance after C-R AEQ}
\begin{figure}[ht!]
\centering
\includegraphics[width=0.65\textwidth,keepaspectratio]{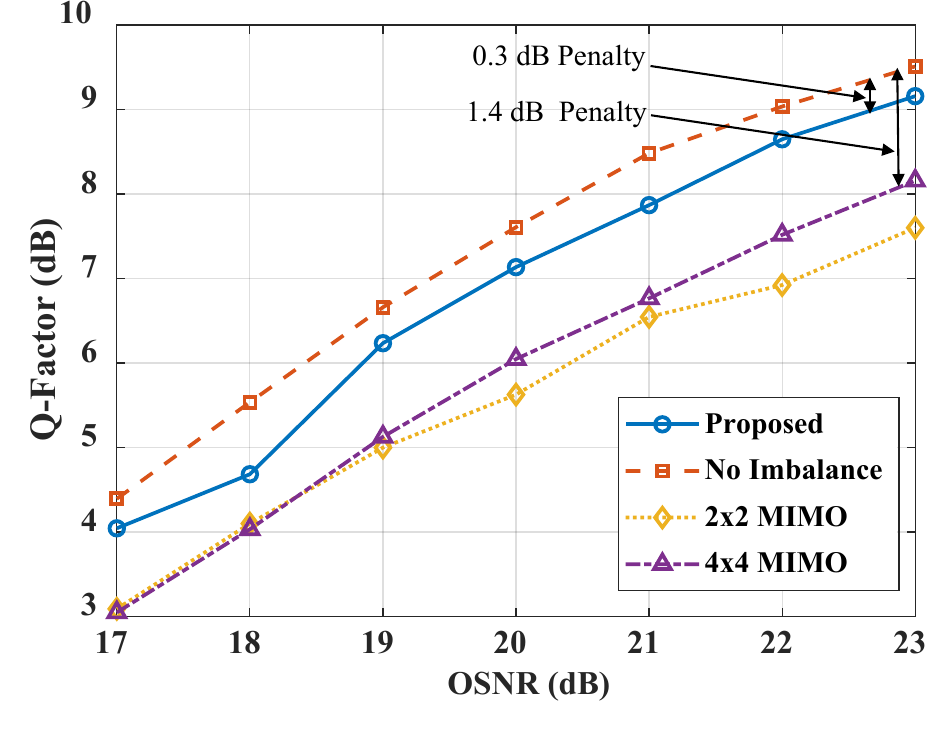}
\captionsetup{justification=justified, singlelinecheck=false}
\caption{Performance of proposed architecture with TX/RX X/Y pol imbalance after 100 km transmission}
\label{Fig.14}
\end{figure}

In real conditions, the IQ imbalance is present in both the transmitter and the receiver. Moreover, we aim to study performance under different OSNR conditions. To simulate that condition, the TX X/Y pol skew is set to 1 ps, since the SDA makes the TX skew less than 1 ps. The TX X/Y pol amplitude imbalance is set to 0.1. The TX X/Y pol phase imbalance is set to 5 degrees. The RX X/Y pol skew is set to 0.25 ps, since the GSMA makes the RX skew less than 0.25 ps. The RX X/Y pol amplitude imbalance is set to 0.1. The RX X/Y pol phase imbalance is set to 5 degrees. These imbalanced values are from our experimental results. The OSNR changes from 17 dB to 23 dB. The equalization performance of our proposed methods is compared against conventional 2×2 and 4×4 MIMO (33 taps per filter), using a perfectly balanced transceiver as the reference. Results are shown in Fig. 14. First, under balanced conditions, the Q-factor varies from 4.39 to 9.5 dB as the OSNR varies from 17 to 23 dB. Second, under imbalanced conditions, the Q-factor changes from 4 to 9.2 dB as the OSNR varies from 17 to 23 dB via our proposed methods. The Q penalty is 0.3 dB when the OSNR changes from 19 dB to 23 dB. For comparison, the 2$\times$2 and 4$\times$4 MIMO Q-factor changes from 3.1 to 7.6 dB and from 3 to 8.1 dB, respectively. The Q penalty is 1.4 dB when the 4$\times$4 MIMO is used at an OSNR of 23 dB. This means that our proposed methods can reduce the Q penalty caused by transceiver IQ imbalance. Comparing our proposed methods with the 4$\times$4 MIMO structure, we observe that the Q factor rises by more than 1 dB at an OSNR of 23 dB.

\subsubsection{Performance with DGD after C-R AEQ}
\begin{figure}[ht!]
\centering
\includegraphics[width=0.65\textwidth,keepaspectratio]{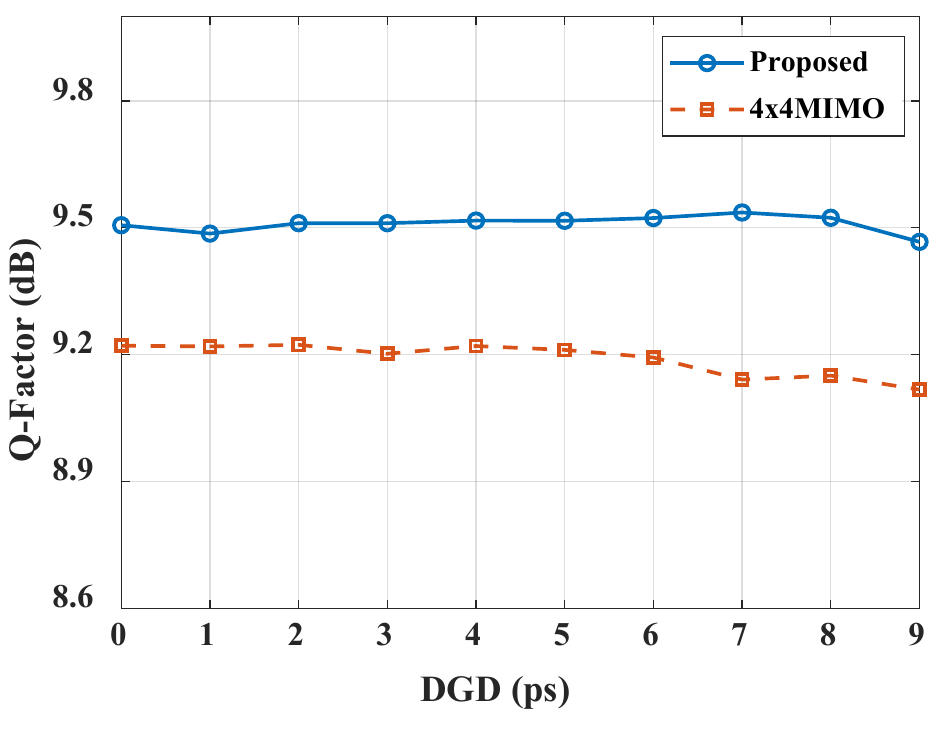} 
\captionsetup{justification=justified,singlelinecheck=false}
\caption{Result of simulated DGD tolerance test}
\label{Fig.15}
\end{figure}
Since the CV-MIMO only has 6 taps per filter, it is necessary to study the performance of PMD compensation. We set the DGD to vary from 0 to 9 ps. All IQ imbalance is set to zero in this part. For comparison, the 4$\times$4 MIMO with 33 taps per filter and the C-R AEQ are used to equalize the signal. As shown in Fig.\ref{Fig.15}, the C-R AEQ is unaffected by DGD values ranging from 0 ps to 9 ps. Moreover, the Q-factor after C-R AEQ is greater than the Q-factor after 4$\times$4 MIMO by about 0.3 dB in the same DGD value.

\section{Experiment}
\begin{figure}[ht!]
\centering
\includegraphics[width=1\textwidth,keepaspectratio]{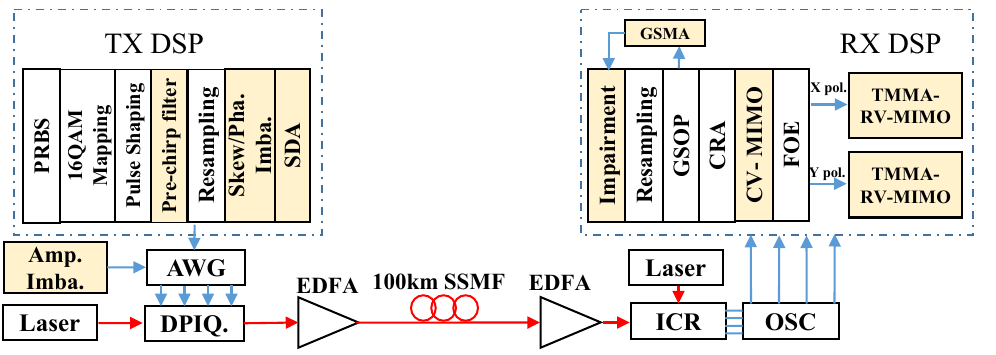} 
\captionsetup{justification=justified,singlelinecheck=false}
\caption{The experimental setup and DSP flow diagram}
\label{Fig.16}
\end{figure}
We investigate the performance of the proposed scheme through experiments. The experimental setup is shown in Fig. \ref{Fig.16}. Two DFB lasers with a linewidth <100 kHz and a wavelength = 1551.12 nm are used as the signal carrier and local oscillator, respectively. A 93.4 GSa/s arbitrary waveform generator (AWG, Keysight M8196A) is used to generate the four tributaries of the 36/50 GBaud DP-Nyquist-16QAM signal with a ROF = 0.5/0.2, and the signal is resampled to 2.5/1.85 SPS. Then the signal is modulated onto the optical carrier using a dual-polarization optical IQ modulator(DPIQ, FTM7977) with a 3-dB bandwidth of 23 GHz. The optical signal is amplified by two EDFAs and transmitted over a 100-km standard single-mode fiber(SSMF). At the receiver, the optical signal is demodulated by an integrated coherent receiver (ICR, CPRV1225A) with a 3-dB bandwidth of 29 GHz. Finally, the signal was sampled by an oscilloscope (OSC, DSOZ594A) at a sample rate of 80 GHz. In TX DSP, the CD is compensated by a pre-chirp filter with a block size of 112. The TX IQ skew/pha. imbalance is added digitally. The TX amp. imbalance is added by AWG. In RX DSP, the signal is resampled to 2 SPS, and the RX IQ imbalance is added digitally first. The signal is input to the CRA after the Gram-Schmidt orthogonalization procedure (GSOP). Meanwhile, the signal is input to the GSMA to calculate the RX skew. To verify the performance of our proposed methods, the experiments were repeated several times. The 36 GBaud signal is used to test the SDA and GSMA performance.

 \begin{figure}[ht!]
\centering
\includegraphics[width=1\textwidth,keepaspectratio]{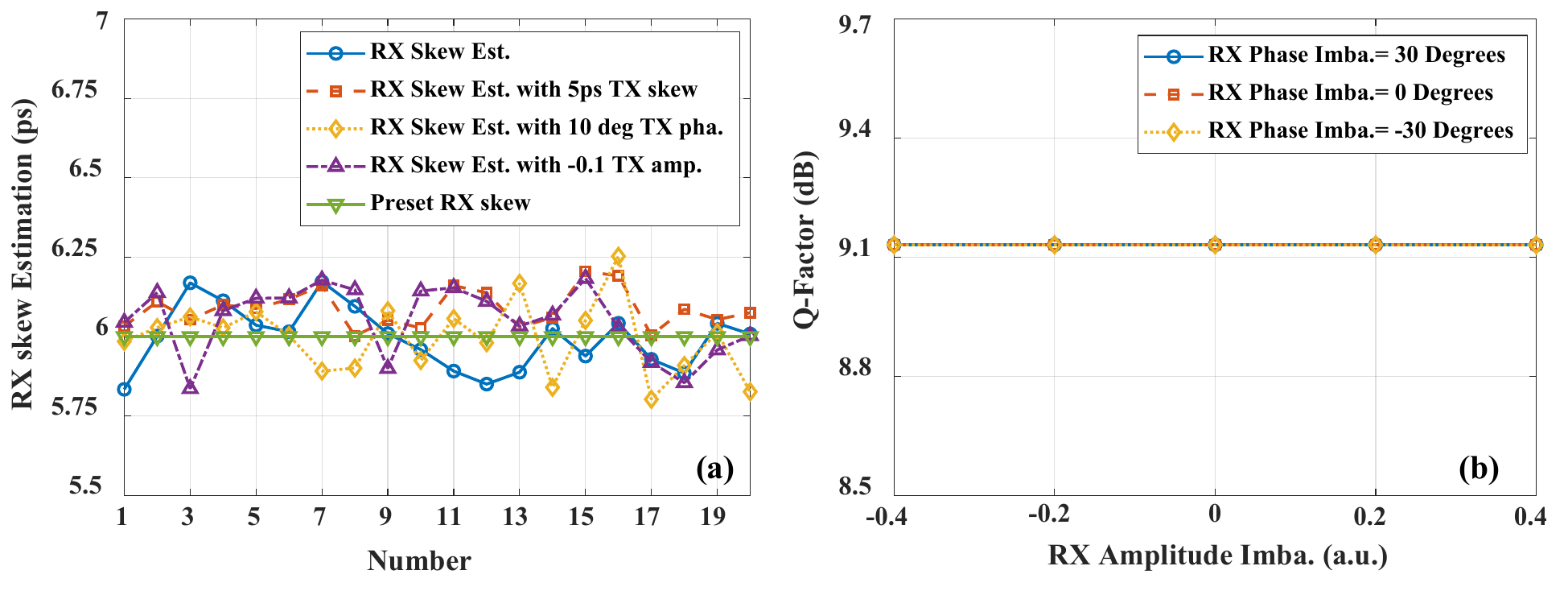} 
\captionsetup{justification=justified,singlelinecheck=false}
\caption{(a)RX skew estimation after 20 times test(b)RX amp./pha. imbalance tolerance of the proposed architecture after 100 km transmission }
\label{Fig.17}
\end{figure}
    Figure \eqref{Fig.17} characterizes the performance of GSMA with the receiver IQ imperfect. The RX skew is set to 6 ps. In Fig. \eqref{Fig.17}(a), the skew estimation is performed 20 times under each kind of IQ imbalance. As shown in Fig. \eqref{Fig.17}(a), most estimation results are around 6 ps, with a minimum of 5.8 ps and a maximum of 6.25 ps. This means the estimation error is less than 0.25 ps. Figure \eqref{Fig.17}(b) displays the relation between Q-factor and receiver IQ amp./pha. imbalance after receiver IQ skew compensation. The Q-factor is unaffected by  IQ amp./pha. imbalance due to the GSOP. 
 \begin{figure}[ht!]
\centering
\includegraphics[width=1\textwidth,keepaspectratio]{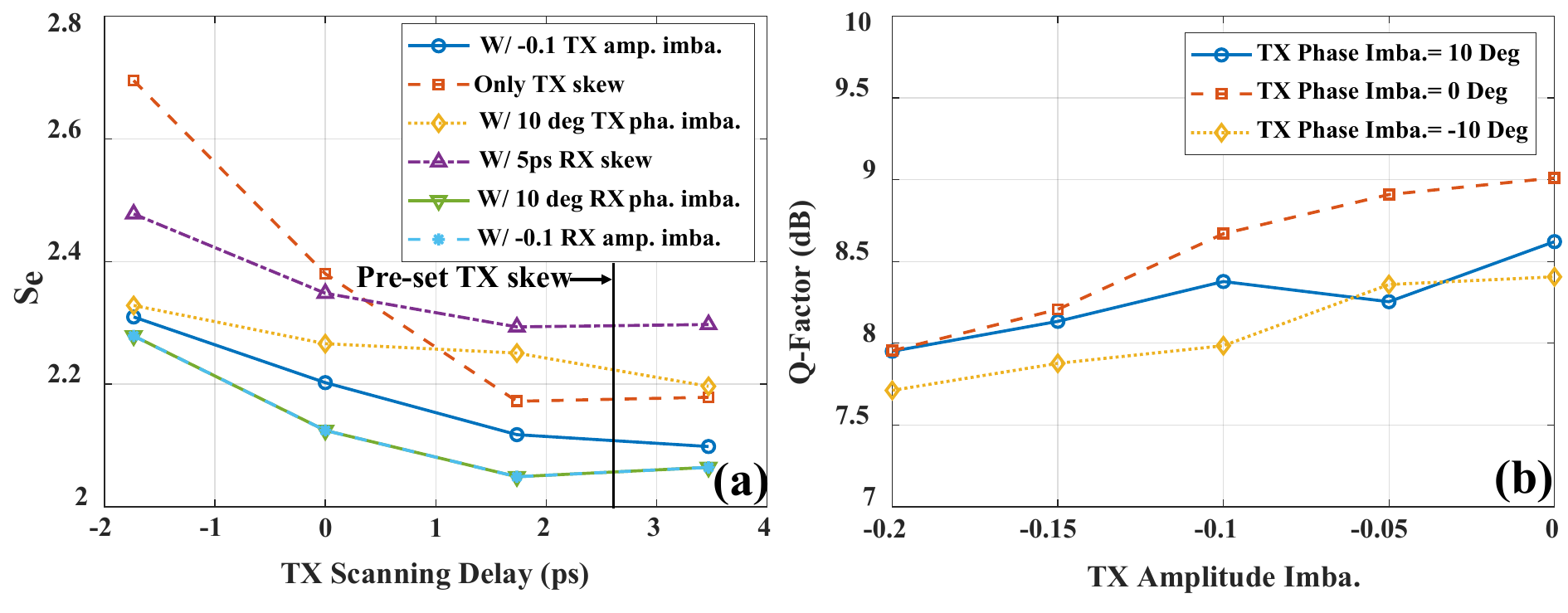} 
\captionsetup{justification=justified,singlelinecheck=false}
\caption{(a)TX skew estimation after 2 times scanning(b)TX amp./pha. imbalance tolerance of the proposed architecture after 100 km transmission}
\label{Fig.18}
\end{figure}

Figure \eqref{Fig.18} characterizes the performance of our proposed methods with the transmitter IQ imperfect after RX X/Y pol skew compensation. In Fig. \eqref{Fig.18}(a), the TX X/Y pol skew is set to 2.6/2.6 ps. The $S_{e}$ is plotted as a function of TX delay. According to Fig. \eqref{Fig.18}(a), the $S_{e}$ is minimized at TX scanning delays of 1.74 or 3.47 ps. We select 2 ps skew as the TX skew estimation result. In  Fig. \eqref{Fig.18}(b), the TX amp./pha. imbalance is studied. The TX X/Y pol amplitude imbalance changes from -0.1 to 0.1, and the phase imbalance is set to -10, 0, and 10 deg respectively. Figure \eqref{Fig.18}(b) shows that the Q-factor penalty is 0.6 dB under a phase imbalance of $\pm$ 10 deg with the -0.1 amplitude imbalance.

\begin{figure}[ht!]
\centering
\includegraphics[width=1\textwidth,keepaspectratio]{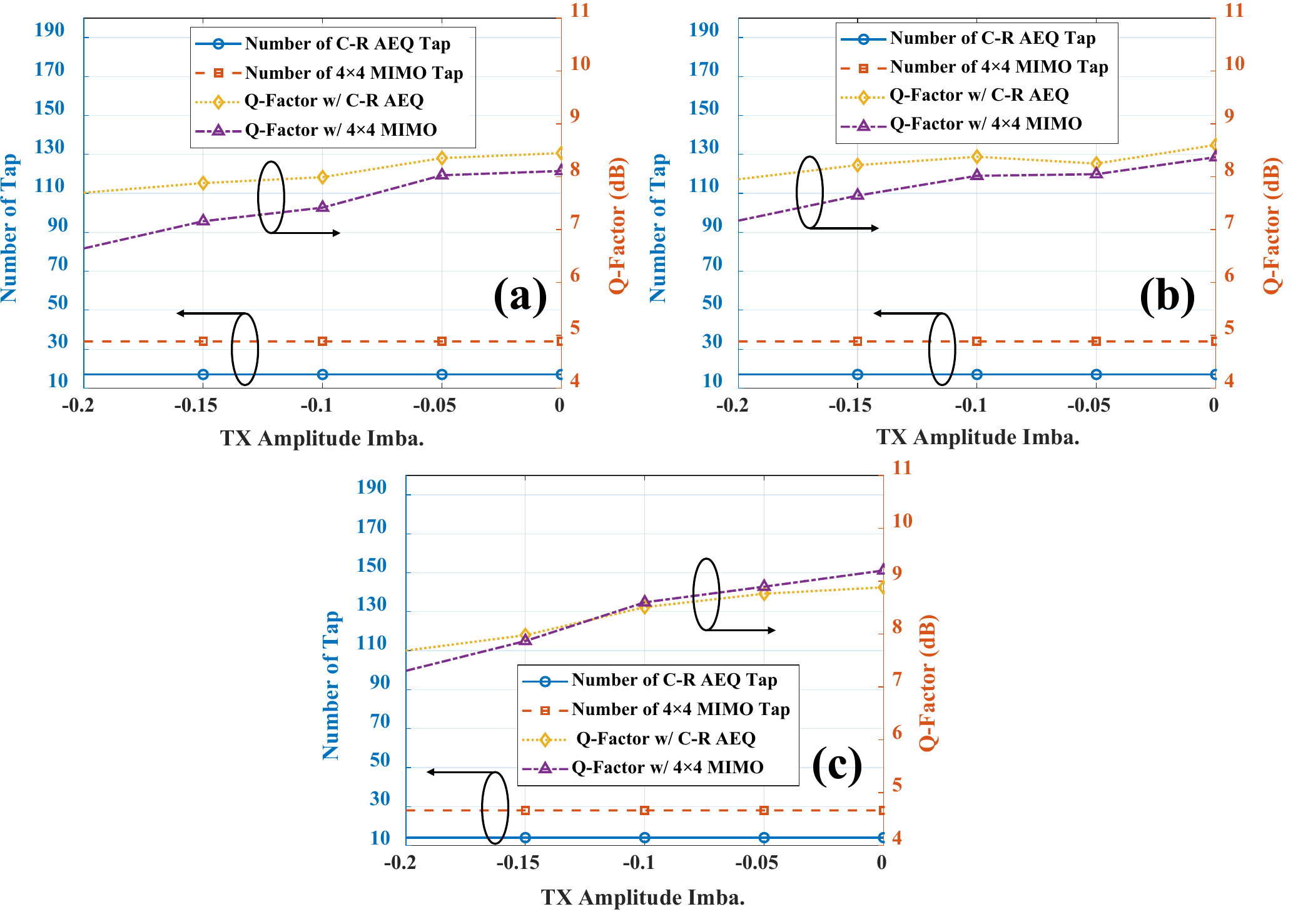} 
\captionsetup{justification=justified,singlelinecheck=false}
\caption{(a)Q-factor under difference taps(b)The C-R AEQ and 4×4 MIMO requared number of taps respectively under the same Q-factor}
\label{Fig.19}
\end{figure}
We compare the performance of the 4×4 MIMO and C-R AEQ as shown in Fig. \eqref {Fig.19}. Because the 4×4 MIMO only compensates for RX X/Y IQ imbalance, for fair comparison, both TX and RX X/Y pol skew are compensated. Moreover, the same received signal is equalized using a 4×4 MIMO or C-R AEQ in each comparison. First, we define the non-butterfly and butterfly structure numbers of filter taps shown in Eq. \ref{eq100} and Eq. \ref{eq101}.
\begin{align}
T_{n}=ceil(T_{pf}*2/4)
\label{eq100}
\end{align}
\begin{align}
T_{b}=T_{pf}*4/4
\label{eq101}
\end{align}
Where $T_{pf}$ is the number of taps per filter, $T_{n}$ is the number of taps in the non-butterfly filter, and $T_{b}$ is the number of taps in the butterfly filter. The CV-MIMO taps are fixed at 6 in C-R AEQ. In Fig.\eqref{Fig.19}(a) or (b), the TX pha. imbalance is -10 deg or 10 deg, respectively. According to Fig. \ref {Fig.19}(a) and (b), the Q-factor using C-R AEQ with 17 taps is higher than the Q-factor using 4×4 MIMO with 34 taps. In Fig.\eqref{Fig.19}(c), the C-R AEQ taps are reduced to 14, and the 4×4 MIMO taps are reduced to 28. It can be observed that the Q-factor penalty is less than 0.2 dB using C-R AEQ under zero TX amplitude imbalance. 

\begin{figure}[ht!]
\centering
\includegraphics[width=1\textwidth,keepaspectratio]{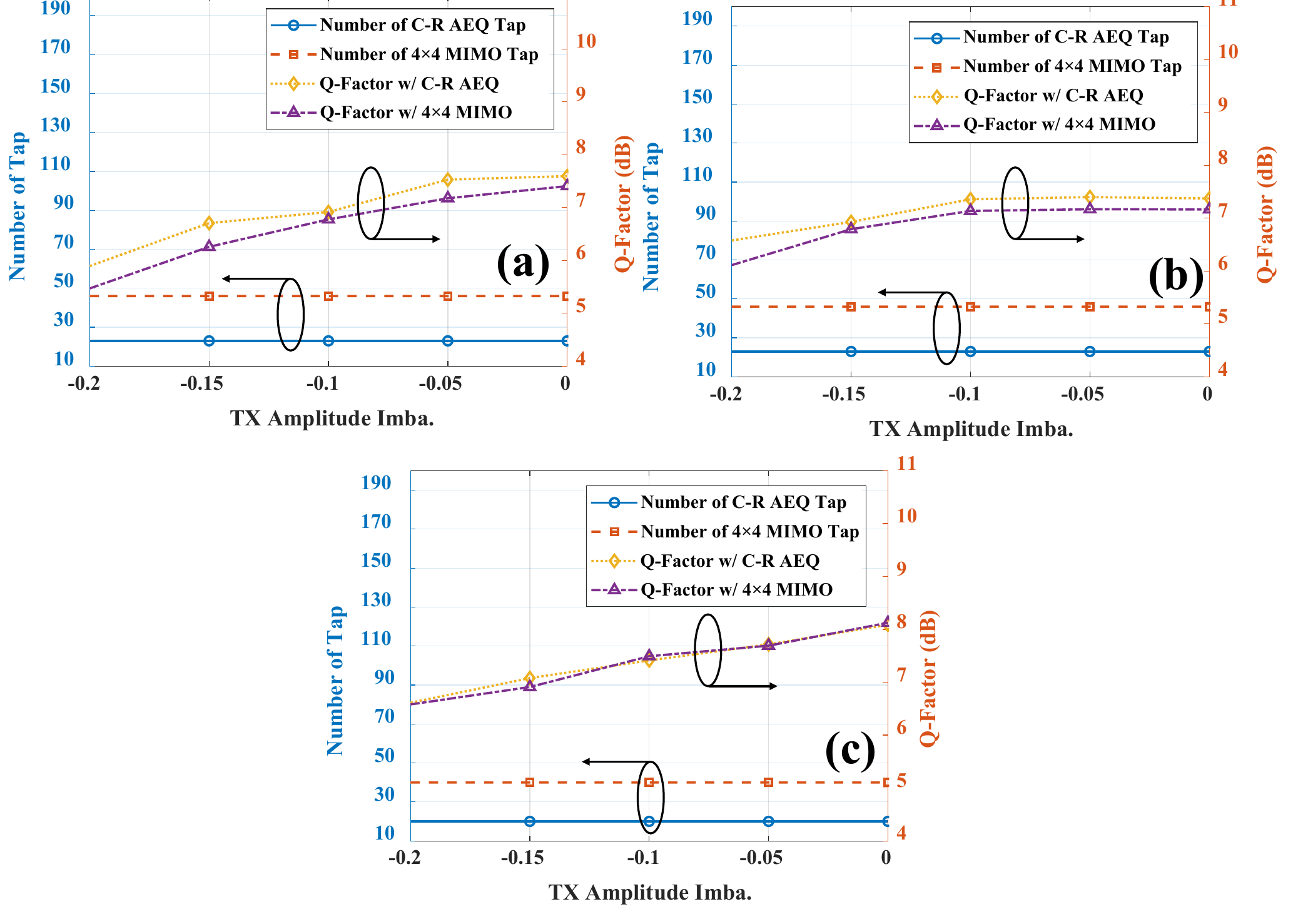} 
\captionsetup{justification=justified,singlelinecheck=false}
\caption{(a)Q-factor under difference taps(b)The C-R AEQ and 4×4 MIMO requared number of taps respectively under the same Q-factor}
\label{Fig.20}
\end{figure}
We also evaluate the performance of C-R AEQ for a 50 GBaud 16QAM system, as shown in Fig.\eqref{Fig.20}. In Fig.\eqref{Fig.20}(a) and (b), the C-R AEQ and the 4×4 MIMO number of taps are set to 23 and 46, respectively. In Fig.\eqref{Fig.20}(c), the C-R AEQ and the 4×4 MIMO number of taps are set to 20 and 40, respectively. As shown in Fig.\eqref{Fig.20}(a) and (b), the C-R AEQ can reduce the Q-factor penalty under TX IQ imbalance. The Fig.\eqref{Fig.20}(c) shows that, at zero IQ imbalance condition, the C-R AEQ needs 20 taps to achieve 8.1 dB Q-factor, and the 4×4 MIMO requires 40 taps to achieve the same Q-factor.

\section{Complexity}
In this part, the complexity of our proposed architecture is studied. The complexity is shown by the number of real multiplications via comparison with other techniques. The signal oversampling rate is twice the symbol rate.
 \subsection{CD Compensation Complexity}
 The pre-chirp filter and FDE are used to compare the complexity. At the signal rate of 36 Gbaud, the block size of the pre-chirp filter is set to 112. According to the data given in reference\cite{ref22}, the pre-chirp filter requires about 14 real multiplications per symbol, whereas the FDE requires about 24 real multiplications per symbol. The complexity is reduced by more than 40$\%$.

 \subsection{IQ Skew Estimation Complexity}

To assess the complexity of the GSMA, we compare GSMA with Godard’s PD-based RX skew estimation \cite{ref18}. According to reference \cite{ref18}, Godard’s PD requires $20(N_{FFT}/4+1)/N_{FFT}+8log_{2}N_{FFT}$ real multiplications per symbol, where $N_{FFT}$ is number of point FFT. Typically, $N_{FFT}$ is set to 128. Thus, the complexity is 61 real multiplications per symbol. In contrast, Gardner’s PD for skew compensation requires 34 real multiplications per symbol, and the detail of complexity is that phase error detection and RX skew compensation require 26 and 8 real multiplications per symbol, respectively. The complexity is reduced by more than 55$\%$. The reason is that the XI, XQ, YI, and YQ all need to perform FFTs in Godard’s PD, which significantly increases complexity.

 Regarding the complexity of SDA, first, the scanning delay block can be implemented by adjusting the transmitter output delay. Second, even without SDA, updating the equalizer coefficients requires equalizer error. Therefore, the TX skew compensation needs approximately zero real multiplications. 
 \subsection{AEQ Complexity}
The AEQ complexity is calculated by comparing the C-R AEQ and 4×4 MIMO AEQ. According to Fig. \ref {Fig.19}(c), the C-R AEQ and 4×4 MIMO AEQ require 14 taps(6-tap CV-MIMO and 16-tap TMMA-RV-MIMO)  and 28 taps, respectively, to achieve the same Q-factor. According to ref.\cite{ref24}, the butterfly filter requires $4\times4\times T_{pf}$ real multiplications per symbol, and the non-butterfly filter requires $2 \times 4 \times T_{pf}$ real multiplications per symbol. Thus, the C-R AEQ requires 224 real multiplications . In contrast, the 4×4 MIMO requires 448 real multiplications. For the 50 Gbaud system, the C-R AEQ requires 320 real multiplications, and the 4×4 MIMO requires 640 real multiplications. In addition, the C-R AEQ can recover the carrier phase with 0 real multiplications, whereas the BPS/ML\cite{CPR} needs 64 real multiplications. Therefore, the C-R AEQ complexity is reduced by 50$\%$.
The numbers of real multiplications per symbol are compared in Table I.

\begin{table}[htbp]
\caption{Real Multiplications Required Per Symbol}
  \label{tab:shape-functions}
  \centering
\begin{tabular}{ccc}
\hline
Procedure& Proposed & Conventional Cases \\
\hline
TX Skew Compensation & 0 & - \\
RX Skew Compensation & 34 & 61\cite{ref18} \\
Equalizer & $224/320$ & 448/640\cite{ref24}\\
CPR& 0 & 64\cite{CPR} \\
\hline
Total& 258/354 & 573/765 \\
\hline
\end{tabular}
\end{table}

\section{Conclusion}
We proposed a low-complexity multi-dimensional IQ imbalance compensation architecture. First, the transceiver IQ skew is estimated and compensated for by GSMA and SDA. Then, the C-R AEQ consists of a CV-MIMO and a TMMA-RV-MIMO equalizer to perform signal equalization, CPR and TX phase/amplitude imbalance compensation. For GSMA, the simulation and experimental results show that the RX skew estimation error is less than 0.25 ps. Compared to Godard’s PD, the complexity of GSMA is reduced by 55\%. For SDA, the simulation and experimental results show that the TX skew estimation error is less than 0.9 ps. For C-R AEQ, it can tolerate $\pm$0.1 TX IQ amplitude imbalance and $\pm$5 degrees with 0.3 dB penalty in the 36 GBaud signal. In 100 km 36 GBaud DP-16QAM signal transmission, compared to 4×4 MIMO, the complexity of C-R AEQ is reduced by more than 50\%.  A key advantage of the proposed GSMA is its hardware reuse. The algorithm directly utilizes the existing clock recovery circuit (CRA), thereby eliminating the need for dedicated circuitry and avoiding additional chip area consumption. Both the SDA and C-R AEQ are low-complexity algorithms as well. Therefore, the architecture can be deployed in hardware-resource limit area.

\bibliography{sample}

\end{document}